\documentclass[]{elsart}

\usepackage{graphicx}

\usepackage{amssymb}

\usepackage{amsmath}

\newcommand \beq{\begin{eqnarray}}
\newcommand \eeq{\end{eqnarray}}
\newcommand \bef{\begin{figure}}
\newcommand \eef{\end{figure}}
\newcommand \bei{\begin{itemize}}
\newcommand \eei{\end{itemize}}
\newcommand \bet{\begin{table}}
\newcommand \eet{\end{table}}

\begin{document}

\begin{frontmatter}

\title{A Hadron Blind Detector for the PHENIX Experiment at RHIC. }
\vspace{-12mm}
\author[Weizmann]{Z. Fraenkel},
\author[Weizmann]{A. Kozlov},
\author[Weizmann]{M. Naglis},
\author[Weizmann]{I. Ravinovich},
\author[Weizmann,budker]{L. Shekhtman},
\author[Weizmann,corr]{I. Tserruya},
\author[BNL]{B. Azmoun},
\author[BNL]{C. Woody},
\author[KEK]{S. Sawada},
\author[RIKEN]{S. Yokkaichi},
\author[SUNY]{A. Milov},
\author[Tokyo]{T. Gunji},
\author[Tokyo]{H. Hamagaki},
\author[Tokyo]{M. Inuzuka},
\author[Tokyo]{T. Isobe},
\author[Tokyo]{Y. Morino},
\author[Tokyo]{S. X. Oda},
\author[Tokyo]{K. Ozawa},
\author[Tokyo]{S. Saito},
\author[Tokyo]{T. Sakaguchi},
\author[Waseda]{Y. Yamaguchi}

\thanks[corr]{Corresponding author.
  Tel.: + 972-8-934 4052; fax: + 972-8-934 6021.
  \it{E-mail address}: Itzhak.Tserruya@weizmann.ac.il
}

\thanks[budker]{on leave from the Budker Institute of Nuclear
  Physics, Novosibirsk 630090, Russia
}

\address[Weizmann]{ Weizmann Institute of Science, Rehovot 76100, Israel }
\address[BNL]{Brookhaven National Laboratory, Upton, NY 11973-5000, USA}
\address[KEK]{KEK, Tsukuba-shi, Ibaraki-ken 305-0801, Japan}
\address[RIKEN] {RIKEN, Wako, Saitama 351-0198, Japan}
\address[SUNY]{Stony Brook University, SUNY, Stony Brook, NY 11794-3400,USA}
\address[Tokyo]{University of Tokyo, Tokyo 113-0033, Japan }
\address[Waseda]{Waseda University, Tokyo 162-0044, Japan}

\maketitle

\begin{abstract}
A Hadron Blind Detector (HBD) is being developed for the PHENIX
experiment at RHIC. It consists of a \v{C}erenkov radiator
operated with pure CF$_4$ directly coupled in a windowless
configuration to a triple-GEM detector element with a CsI
photocathode and pad readout. The HBD operates in the bandwidth
6-11.5 eV (110-200~nm). We studied the detector response to
minimum ionizing particles and to electrons. We present
measurements of the CsI quantum efficiency, which are in very good
agreement with previously published results over the bandwidth
6-8.3 eV and extend them up to 10.3 eV. Discharge probability and
aging studies of the GEMs and the CsI photocathode in pure CF$_4$
are presented.
\end{abstract}
\begin{keyword} HBD \sep GEM \sep CsI photocathode \sep UV-photon detector \sep
CF$_4$
\PACS 29.40.-n \sep 29.40.Cs \sep 29.40.Ka \sep 25.75.-q
\end{keyword}

\end{frontmatter}

\section{Introduction}

 We describe the operation and performance of a Hadron Blind Detector (HBD) which is being
developed as an upgrade of the PHENIX detector at the Relativistic
Heavy Ion Collider (RHIC) at BNL. The main purpose of the HBD is
to allow the measurement of low-mass ($m_{e^+e^-} \leq$ 1
GeV/c$^2$) electron-positron pairs produced in central heavy ions
collisions at energies up to $\sqrt{s_{NN}}$ = 200 GeV.
  Low-mass dileptons are a powerful observable in the quest for the quark-gluon plasma and in
particular for the restoration of chiral symmetry expected to take
place in ultra-relativistic nuclear collisions. The results of the
CERES experiment~\cite{ceres} at CERN support the unique physics
potential of this probe. The strong enhancement of low-mass pairs
observed in nuclear collisions could only be explained by invoking
the $\pi\pi$ annihilation channel $\pi^+\pi^- \rightarrow \rho
\rightarrow e^+e^-$ with an in-medium modification of the
intermediate $\rho$ meson which could be linked to the restoration
of chiral symmetry~\cite{rapp-brown}.

With its excellent mass resolution, the PHENIX detector has the potential to perform
precision spectroscopy of the $\rho$, $\omega$, and $\phi$ mesons in addition to the pair
continuum measurement. The observation of mass shifts of the $\rho, \omega$, and $\phi$ mesons
would provide direct evidence for the scenarios invoking the restoration of chiral symmetry.
However, the present configuration of the PHENIX  detector is severely limited in the $e^+e^-$
low-mass region. The detector lacks the capability to recognize and reject
the overwhelming yield of combinatorial background pairs, i.e.
uncorrelated pairs formed by tracks from unrecognized conversions and
$\pi^0$ Dalitz decays. In the mass range $m_{e^+e^-} \sim$ 0.3 - 0.5~GeV/c$^2$ the present
signal to background ratio is  S/B $\sim$ 1/300, making the measurement of the low-mass pair
continuum practically impossible.

  An upgrade of the PHENIX detector is therefore necessary for this measurement.
The present work is focussed on the development of an HBD which is
the key element for such an upgrade. The HBD consists of a
\v{C}erenkov radiator, operated with pure CF$_4$ and directly
coupled to a triple Gas Electron Multiplier (GEM) \cite{Sauli_GEM}
photon detector element \cite{NIM1,dirk1}. The concept is
described in detail in Section 2.

In a recent publication \cite{NIM1} we have demonstrated the successful operation of a
triple-GEM detector in pure CF$_4$. In particular we have shown
that the detector operates in a stable manner at gains up to
10$^4$ with and without a reflective CsI photocathode evaporated
on the upper face of the top GEM. In the present paper, we
concentrate on the hadron blindness properties and performance of
the HBD. We present results obtained with a Hg UV lamp, a
$^{55}$Fe X-ray source and an $^{241}$Am alpha source. We also
present results from beam tests carried out at KEK using 1 GeV/c
pions. Section 3 describes the various
setups and conditions under which the measurements were performed.
Section 4 presents measurements of the detector response to
photoelectrons, alpha particles and pions as a function of the
drift field. In Section 5,  we present our measurements of the CsI
quantum efficiency over the bandwidth 6-10.3 eV.
The discharge probabilities of HV-segmented GEM foils are
discussed in Section 6 and aging studies of the GEM foils and the
CsI photocathode in pure CF$_4$ atmosphere are presented in
Section 7. A short summary and conclusions are presented at the
end of the paper, in Section 8.

\section{The HBD concept. }

\begin{figure}[ht]
\centering
\includegraphics[keepaspectratio=true, width=10cm]{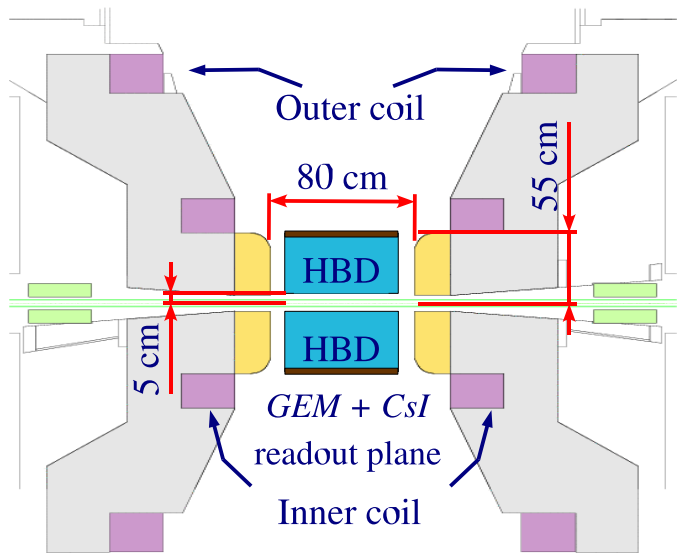}
\caption {Layout of the inner part of the PHENIX detector showing the location of the
          HBD and the second coil.}
   \label{fig:hbdinphnx}
\end{figure}

The PHENIX detector was designed anticipating that the measurement
of low-mass pairs would be feasible with an appropriate upgrade.
In particular, provision was made for the installation of an inner
coil which would create an almost field-free region close to the
vertex, extending out to $\sim$50-60 cm in the radial direction.
In addition to this coil, the second major element of the upgrade
is an HBD located in this field free region. The main task of the
HBD is to recognize and reject conversions and $\pi^o$ Dalitz
decay pairs by exploiting their small opening angle. Its size is
constrained by the available space starting outside the beam pipe
(at r$\sim$5 cm) and ending before the inner coil (at
r$\sim$55 cm). Fig.~\ref{fig:hbdinphnx} shows the layout of the
inner part of the PHENIX detector together with the location of
the coils and the proposed HBD.

The  system specifications of the HBD have been extensively
studied in Monte Carlo simulations. \cite{TechnicalNote391}. A
reduction of the combinatorial background originating from
conversions and $\pi^0$ Dalitz decays of at least two orders of
magnitude can be achieved with a detector that provides electron
identification with a very high efficiency, of at least 90\%. This
also implies a double (electron) hit recognition at a comparable
level. On the other hand, a moderate $\pi$ rejection factor of
$\sim$ 100 is sufficient. Possible realizations of the HBD
detector were considered in the study. After careful consideration
of relevant options for the key elements (gases, detector
configuration and readout chambers), the choice that emerged is a
windowless \v{C}erenkov detector, operated with pure CF$_4$ in a
proximity focus configuration, with a CsI photocathode and
a triple-GEM detector element with pad readout.

   Since a mirror-type RICH detector in the center of PHENIX is very difficult or nearly
impossible to implement, we consider a scheme without  mirror and
without window in which the \v{C}erenkov light from particles
passing through the radiator is directly collected on a CsI
photosensitive cathode plane, forming a circular blob image rather
than a ring as in a RICH detector. The choice of CF$_4$ both as
radiator and detector gas in a windowless geometry results in a
very large bandwidth (from $\sim$6 eV given by the threshold of
the CsI to $\sim$11.5 eV given by the CF$_4$ cut-off) and
consequently in a very large figure of merit N$_0$ and a very
large number of photoelectrons N$_{pe}$ that was estimated to be
of the order of 40 in a 50 cm long radiator
\cite{TechnicalNote391}. This large number of photoelectrons
ensures a very high electron efficiency, and more importantly, it
is crucial for achieving a double-hit recognition larger than
90\%.

   Another important advantage of the present design using GEMs is that it allows the use
of a reflective photocathode (i.e. the top face of the first GEM
is coated with a thin layer of CsI and the photoelectrons are
pulled into the holes of the GEM by their strong electric field)
and consequently the photocathode is totally screened from photons
produced in the avalanche process.

  The present readout scheme foresees the detection of the \v{C}erenkov photoelectrons in
a pad plane with hexagonal pads of size slightly smaller than the
blob size ($\sim$10 cm$^2$) such that the probability of a single
pad hit by an electron entering the HBD is negligibly small. This
results in a low granularity detector. In addition, since the
photoelectrons produced by a single electron will be distributed
between at most three pads, one can expect a primary charge of at
least 10 electrons/pad, allowing operation of the detector at a
relatively moderate gain of a few times 10$^3$.

\section{Setup and experimental conditions. }

\begin{figure}[ht]
    \vspace{-5mm}
    \centering
    \includegraphics[keepaspectratio=true, width=14cm]{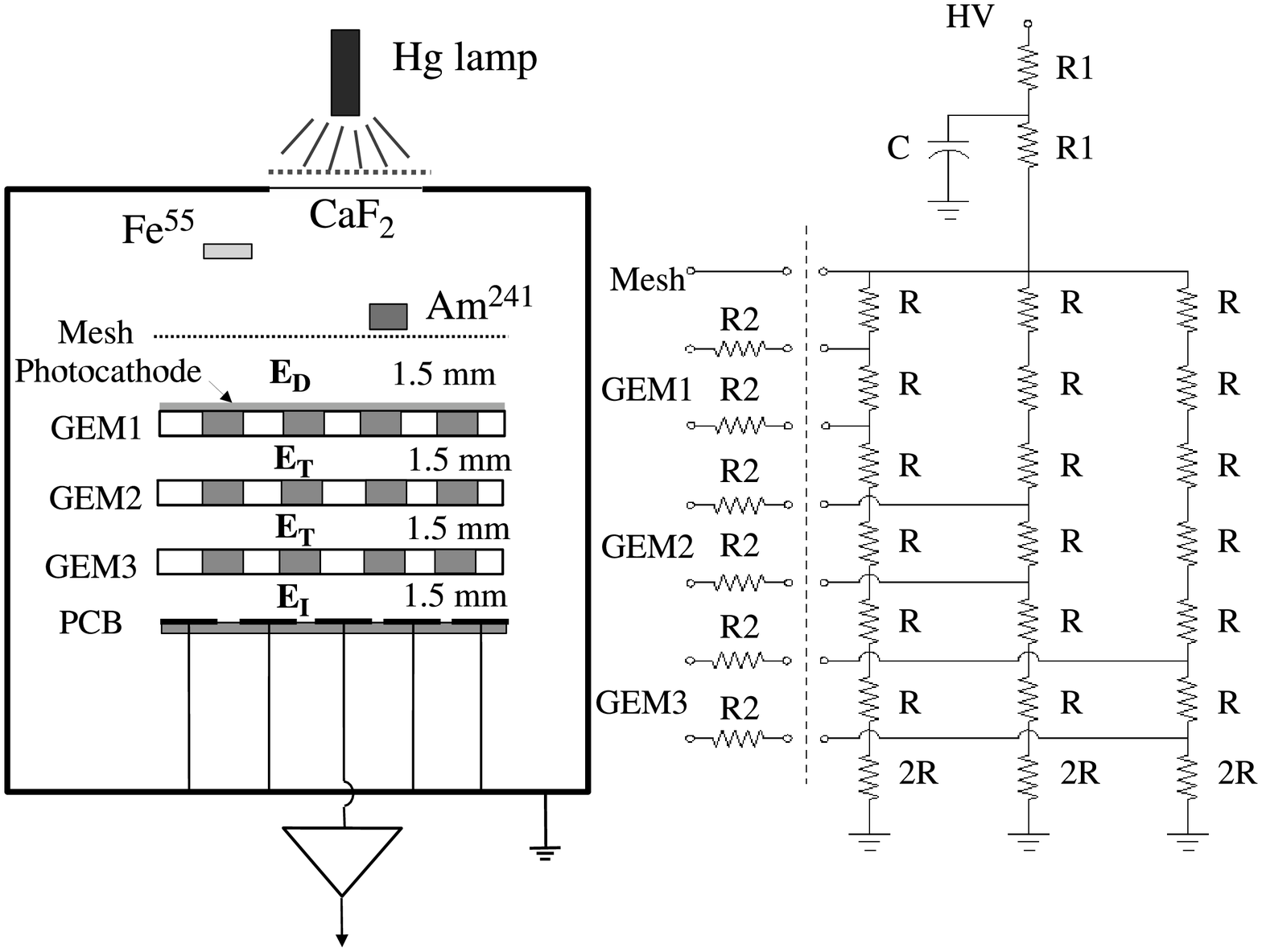}
    \vspace{-5mm}
    \caption{Setup of the triple-GEM detector and resistor chain. The Hg lamp,
    $^{55}$Fe and $^{241}$Am sources were used for measurements with UV-photons,
    X-rays and $\alpha$-particles, respectively.}
    \label{fig:assembly}
\end{figure}

 GEMs produced at CERN were used in all measurements. The GEMs were
 made of  50~$\mu$m kapton foils with 5~$\mu$m thick copper layers,
 60-80~$\mu$m diameter holes and 140~$\mu$m pitch. They had a
 sensitive area of 30$\times$30 mm$^2$  or 100$\times$100 mm$^2$. These two types of
 GEMs will be referred to in the text as "small" and "large" respectively. The top side
 of some of the large GEMs was divided into four segments (25$\times$100 mm$^2$ each)
and will be referred to in the
 text as "segmented". 3 GEMs were assembled into one stack with G10 frames as shown in
 Fig.~\ref{fig:assembly}. The distance between the drift mesh and the top GEM (GEM1), the distance
 between GEMs and the distance between the bottom GEM (GEM3)
 and the printed circuit board (PCB) were all equal to 1.5 mm.
The PCB consisted of 9 square pads of size 33$\times$33 mm$^2$.
Each pad was connected to a charge sensitive preamplifier. CF$_4$
with 99.999\% purity was used in all measurements.

In most tests, the high voltage to the GEMs and mesh was supplied
by a single HV power supply through a three-branch resistive chain
(Fig.~\ref{fig:assembly}). In this scheme three independent
resistive dividers provide the voltage to the top and bottom
planes of each GEM. When a discharge occurs in one GEM  the
voltage across the other GEMs drops only slightly unlike the case
of a single-branch resistive divider. In the latter case the
voltages across other GEMs increase and can cause multiple
discharges and damage the GEMs. The resistors R in the chain were
5.6 M$\Omega$, R1 and R2 were  1.2~M$\Omega$. For the segmented
GEMs, the resistors R2 were connected directly to each segment and
were located inside the detector box. To ensure almost 100\%
electron collection efficiency to the PCB, the induction field
E$_I$ is set to twice the transfer field E$_T$ \cite{NIM1}. For
the measurements where the drift field was varied, two HV power
supplies were used, one for the mesh and one for the three GEMs.

For the measurements with the reflective CsI photocathode GEM1 was
prepared with a gold-coated surface to prevent chemical
interaction with cupper. Below the gold layer a thin Ni layer was
deposited in order to provide good adhesion of the gold layer. The
gold-coated GEMs were also produced at CERN and the deposition of
the CsI layer was performed at the Radiation Detection Physics
Laboratory at the Weizmann Institute. The thickness of the CsI
layer was kept to $\sim$ 2000~$\rm{\AA}$. For the operation with
the reflective photocathode the drift field is set to be equal to
zero or even slightly reversed in order to collect only the
photo-electrons from the CsI layer and to repel the ionization of
minimum ionizing particles. For these measurements the
corresponding contact of the resistive chain was disconnected and
an independent power supply was used for the mesh.

The GEMs, PCB and drift mesh were mounted inside a stainless steel
box which could be pumped down to 10$^{-6}$ Torr and was connected
to inlet and outlet gas lines to allow gas flushing. The setup
contained also devices for the precise measurements of
temperature, pressure and water content down to the ppm level. For
the tests with $^{55}$Fe X-rays, the  radioactive source was
positioned on a moving arm inside the box at a distance of
$\sim$~40 mm from the mesh. The $^{55}$Fe source could be moved
out of the sensitive volume. The total rate of X-rays was kept at
the level of 1 kHz. The 5.9~keV photons from $^{55}$Fe release 110
electrons in CF$_4$ (54~eV per electron-ion pair). For the
measurements of the discharge limits with heavily ionizing
particles we used an $^{241}$Am source which emits 5.5~MeV
alpha-particles. The source was attached to a moving arm that
could be inserted at a distance of $\sim$ 1mm from the mesh and
was strongly collimated in order to provide high energy deposition
and small energy dispersion in the sensitive gap. The rate of
alpha-particles was $\sim$ 100 Hz.

The measurements with the CsI reflective photocathode were
performed with a Hg-lamp through a UV-transparent CaF$_2$ window
mounted on the cover of the detector box. The lamp was positioned
above the window with an absorber which reduced the UV flux by
about a factor of 1000 in order to prevent damage to the
photocathode. The illuminated area of the detector was about 1
cm$^2$. In this geometry the measured photo-electron current was
about 2$\cdot$10$^6$ electrons/mm$^2\cdot$s.

The study of the gain limits required a reliable way to monitor
discharges in the triple-GEM assembly  via the resistive chain.
The chain was biased by a CAEN-126 HV-power supply. This module
includes protection against over-current with a precision of
0.1~$\mu$A. The current in the resistive chain was in the range
between 250 and 300~$\mu$A and the protection threshold was always
kept at 1~$\mu$A above the normal value. This was enough to cause
a trip when a discharge occurred in a GEM. The trip signal was
reset after 1~second and was also counted by a scaler.

The beam test at KEK was done with a 1~GeV/c secondary beam of
negative particles (mainly pions) containing a few percent of
electrons. The setup consisted of two gas \v{C}erenkov counters
(GCC), a set of scintillation counters (S1, S2 and S3), the  HBD
and a lead glass calorimeter (PbGl). The HBD consisted of a 50 cm
long radiator directly coupled to the detector box and setup
described above. The detector was operated with pure CF$_4$ at a
relatively high rate of 100~Hz/cm$^2$ and at a gain of
$\sim$~$10^4$. S1 (100$\times$45~mm$^2$) was in front of the two
GCC's, S2 (25$\times$10~mm$^2$) was just in front of the HBD and
S3 (50$\times$45~mm$^2$) was behind the HBD and in front of the
PbGl calorimeter. The trigger was defined by a coincidence between
the three scintillation counters S1$\cdot$S2$\cdot$S3. Pions and
electrons were selected offline using the data from the two GCCs,
the PbGl and the time-of-flight measured between S3 and S1.

\section{Detector response as function of the drift field.}

 A hadron blind detector is characterized by its insensitivity to
 hadrons, i.e. by a large
hadron rejection factor while keeping a high detection efficiency
for electrons. The hadron blindness property of the proposed
detector is achieved by reversing the direction of the drift field
E$_D$ thereby pushing most of the ionization charges towards the
mesh. With this negative drift-field configuration, photoelectrons
released from the CsI photocathode surface are still effectively
collected into the GEM holes due to the strong electric field
inside the holes which is typically of the order of 100~kV/cm.

In order to characterize and quantify both the hadron rejection
factor and the photo-electron detection efficiency, we performed
systematic measurements as a function of the drift field with
alpha particles, pions of 1 GeV/c and UV photons. Details of these
measurements and the results are presented in this section.

\subsection {Detector response to minimum ionizing particles and alpha particles.}

\begin{figure}[ht]
    \centering
    \includegraphics[keepaspectratio=true, width = 14cm]{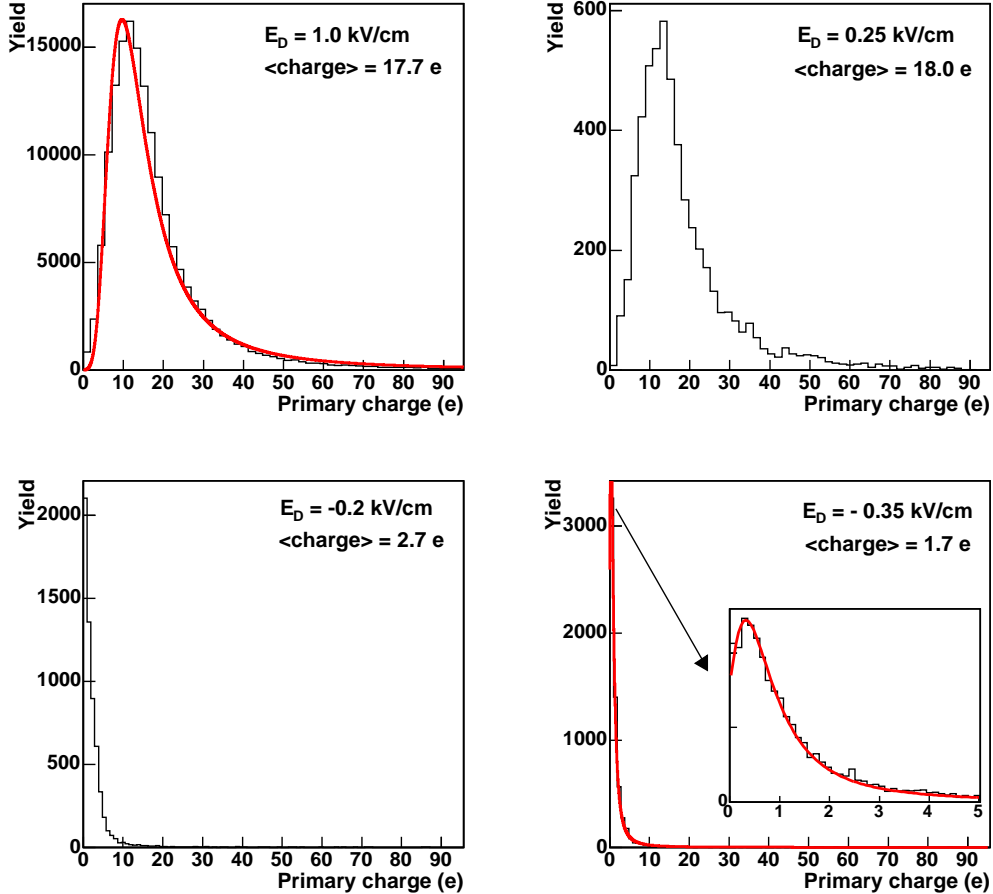}
    \caption{Pulse-height spectra measured with 1~GeV/c pions at various values of the drift field
          E$_D$ in the gap between the mesh and the upper GEM.  The solid lines in the upper
          left and bottom right panels represent fits to a Landau distribution of
          the measured spectra. The insert in the bottom right panel is an expanded view of the
          low signal part of that panel.}
    \label{fig:spectra}
\end{figure}

Fig.~\ref{fig:spectra} shows the pulse-height distribution, after
pedestal subtraction, measured at KEK with 1~GeV/c pions for
various values of E$_D$. The signal is expressed in terms of the
primary ionization charge, using the $^{55}$Fe spectrum measured
under identical conditions. For E$_D$ = +1~kV/cm, the measured
mean amplitude is $\sim$ 18 e corresponding to a primary
ionization of 120~charges/cm or  54~eV/ion-pair assuming an
energy loss of dE/dx = 7~keV/cm \cite{CF4energyloss}. The spectrum
is well reproduced by a Landau distribution characteristic of the
energy loss of a minimum ionizing particle (mip).

The spectrum remains practically unchanged as long as E$_D$ is
positive as shown for two cases in the upper panels of
Fig.~\ref{fig:spectra}. As soon as E$_D$ is reversed, i.e set to
negative values, there is a sharp drop in the pulse height as the
primary charges get repelled towards the mesh. The mean amplitude
drops to $\sim$~10\% of its value for a positive field. This value
results from the collection of ionization charges (i) from a thin
layer above the first GEM surface and (ii) from the entire first
transfer gap which are subject to a two-stage amplification. At a
gain of 10$^4$, the former is estimated to be a factor of $\sim$2
larger than the latter, indicating that when the drift is
reversed, ionization charges are collected from a layer of $\sim
100 \mu$ above the first GEM.

\begin{figure}[ht]
    \centering
    \includegraphics[keepaspectratio=true, width = 12cm]{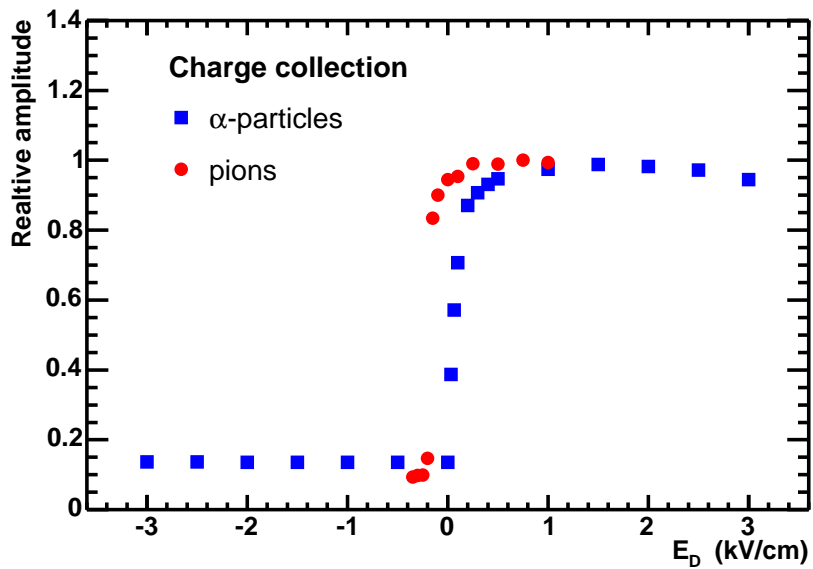}
    \caption{Collection of ionization charge vs. the drift field E$_D$ in the gap between the
      mesh and the upper GEM.}
    \label{fig:blindness}
\end{figure}

The mean amplitude vs. E$_D$ is shown in Fig.~\ref{fig:blindness}.
The amplitude decreases sharply when the polarity of the drift
field is switched and this occurs within a $\Delta$E$_D$ range of
$\sim$0.1~kV/cm. The figure shows also the results of similar
measurements performed with alpha particles. The results are
practically identical in both cases. The small difference in the
values of the field at the onset of the signal drop is well within
the uncertainties of the absolute high-voltage values given the
power supplies used in the two measurements.

\begin{figure}[ht]
    \centering
    \includegraphics[keepaspectratio=true, width = 12cm]{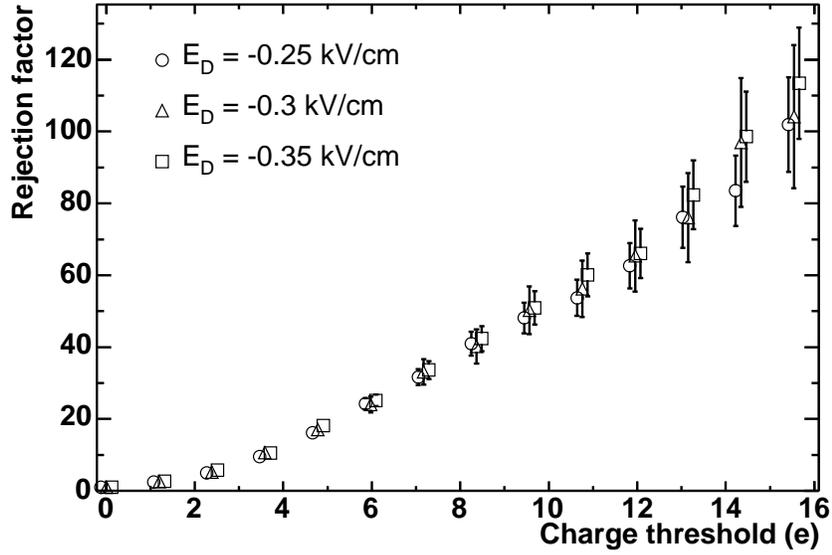}
    \caption{Hadron rejection factor derived from the pion pulse-height distribution as a function
          of the amplitude threshold in units of the primary
          ionization charge. The error bars represent the statistical uncertainties.}
    \label{fig:rejf}
\end{figure}

The hadron rejection factor derived from the pion spectra measured
at negative drift fields is shown in Fig.~\ref{fig:rejf}. The
rejection is limited by the long Landau tail and depends on the
amplitude threshold that can safely be applied without
compromising the electron collection efficiency. Rejection factors
of the order of 50 can be achieved with an amplitude threshold of
$\sim$ 10~e. A much higher rejection factor is achieved by
combining the amplitude response with the hit size. As mentioned
above, the pad readout consists of hexagonal pads with a size
somewhat smaller than the Cherenkov blob size. Under these
conditions charged particles will produce single pad hits whereas
electrons will most probably produce multiple-pad hits thereby
providing an additional powerful handle on the charged particle
rejection.

\subsection {Detector response to photoelectrons.}

\begin{figure}[ht]
    \centering
    \includegraphics[keepaspectratio=true, width = 12cm]{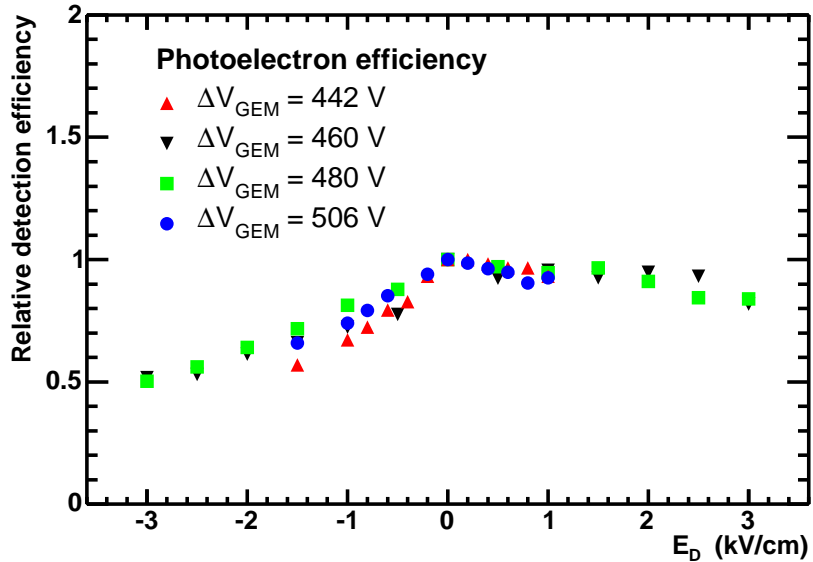}
    \caption{The photoelectron detection efficiencies for different gains vs. the electric
    field E$_D$ in the gap between the mesh and the upper GEM.}
    \label{fig:photodetection}
\end{figure}

Measurements as a function of the drift field to determine the
electron detection efficiency were performed with UV-photons from
a Hg lamp irradiating the CsI photocathode through the
UV-transparent CaF$_2$ window. We measured the photo-current at
the PCB for values of the HV accross the GEMs  varying from 442 to
506~V which correspond to gas gain variations of more than a
factor of $\sim$40. The results are shown in
Fig.~\ref{fig:photodetection}. The various measurements have been
normalized to 1 at E$_D$=0~kV/cm to represent the relative
detection efficiency and to allow an easy comparison between the
measurements. The relative detection efficiency is practically
independent of the field

across the GEMs. The efficiency slightly increases as the positive
drift field decreases, reaching a maximum at 0~kV/cm. A mild
decrease in the current is then observed as the drift field
becomes more and more negative demonstrating that the detection
efficiency of the photoelectrons remains very high even at
negative drift fields of 1~kV/cm. Combining the results of
Figs.~\ref{fig:blindness} and \ref{fig:photodetection} one sees
that the best performance is achieved by applying a slightly
negative field in the drift gap. The results presented here are
consistent with those of ref. \cite{mormann,qe11} though with
different gases and voltages across the GEMs.


\section {CsI Quantum Efficiency}

The absolute quantum efficiency (QE) of the CsI photocathode was
measured by a large number of groups (see~\cite{qe11}, \cite{qe13}
for comprehensive reviews and further references).
 Most of the measurements are in reasonable agreement with each other.
However, none of these measurements were performed at wavelengths
below 150~nm (or a photon energy above 8.3~eV). Since CF$_4$ is
transparent up to 11.5~eV it was important  to extend the
measurements of the  absolute QE of CsI as much as  possible.

\begin{figure}[ht]
    \centering
    \includegraphics[keepaspectratio=true, width = 115mm]{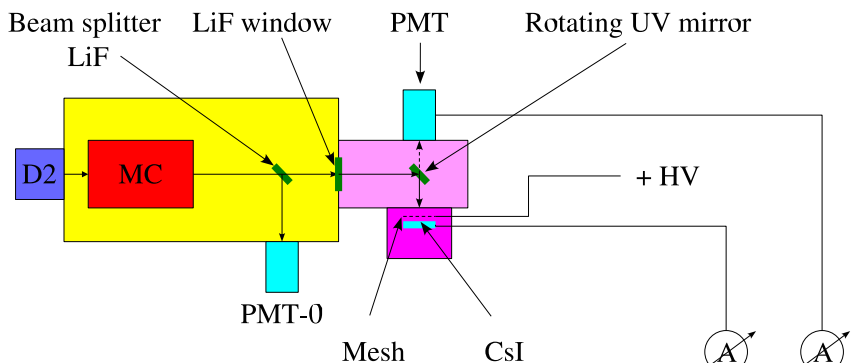}
\vspace{-5mm}
\caption{Schematic view of the experimental setup for measuring the
  quantum efficiency of the CsI layer.}
\label{fig:setup}
\end{figure}

The determination of the absolute QE requires an absolutely
calibrated light source, which is not available in most laboratories. Therefore, usually
a relative method is used, namely the recording of the sample response relative
to a "known" reference.
In our measurements we used as a reference a calibrated
photomultiplier tube.

The experimental setup used for the determination of the absolute
quantum efficiency of the CsI layer is shown in
Fig.~\ref{fig:setup}. It includes a vacuum ultraviolet (VUV)
monochromator (Jobin Yvon H20, 115-500~nm) equipped with a deuterium lamp
(Hamamatsu L7293, 115-320~nm), coupled via a LiF window (cut-off at 105~nm)
to a detector box. The monochromator box also includes a LiF beam
splitter which splits the beam between the photomultiplier PMT-0 (Hamamatsu R1460)
and the detector box. PMT-0 serves as a normalization to monitor
the deuterium lamp intensity. The detector box included an
absolutely calibrated photomultiplier tube PMT (Hamamatsu R6836)
on one side and on the other side a  box containing a small
(3$\times$3~cm$^2$) GEM foil on which a 2500~$\rm{\AA}$  layer of
CsI was evaporated. PMT was operated in photodiode mode (gain=1).
Above the foil and at a distance of 1.5~mm
from it was a mesh electrode which was at a positive voltage with
respect to the foil. The detector box also had a UV mirror which
served to deflect the beam alternatively to the CsI layer and to
the PMT. Collimators of 8~mm diameter were placed in front of the
mesh and PMT, making sure that the solid angle seen by the
photomultiplier and the CsI layer was exactly the same. By
rotating the UV-mirror the current was measured in turn over the
whole wavelength range on both devices.

\begin{figure}[ht]
    \centering
    \includegraphics[keepaspectratio=true, width = 12cm]{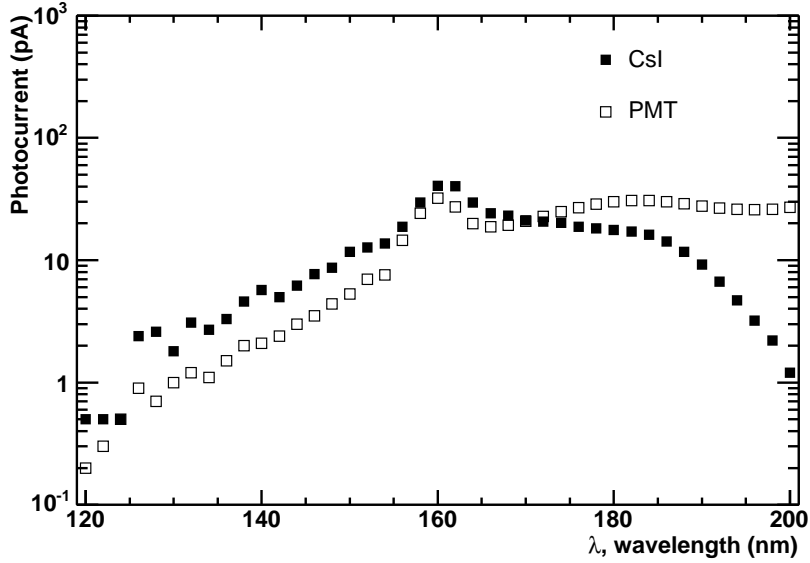}
    \caption{Photocurent from the CsI layer and the reference photomultiplier
             PMT as function of wavelength.}
    \label{fig:photocurrent}
\end{figure}

The current of the CsI and photomultiplier PMT as measured in
vacuum is shown in Fig.~\ref{fig:photocurrent}. The measurements
were done over the wavelength range of 120 - 200~nm (E = 6.2 -
10.3~eV). The measurements were repeated with $CF_{4}$ gas at
atmospheric pressure. The total path in CF$_4$ was 23~cm.

The absolute quantum efficiency of the CsI layer at a given
wavelength $\lambda$ is given by:

\beq
(QE)_{CsI}(\lambda) = \frac{(QE)_{PMT}(\lambda) * I_{CsI}(\lambda)}{I_{PMT}(\lambda) * C_{1} * C_{2}}
\label{eq:quantum-efficiency}
\eeq

where $(QE)_{PMT}(\lambda)$ is the absolute quantum efficiency of
the PMT at the wavelength $\lambda$, $I_{CsI}(\lambda)$ - the CsI
photocathode current measured at that wavelength, $I_{PMT}$ - the
PMT photomultiplier current at $\lambda$ , $C_{1}$ -  the mesh
transparency ($C_{1}$ = 0.81), and $C_{2}$ - the opacity of the
CsI layer due to the GEM holes ($C_{2}$ = 0.833).

\begin{figure}[ht]
    \centering
    \includegraphics[keepaspectratio=true, width = 12cm]{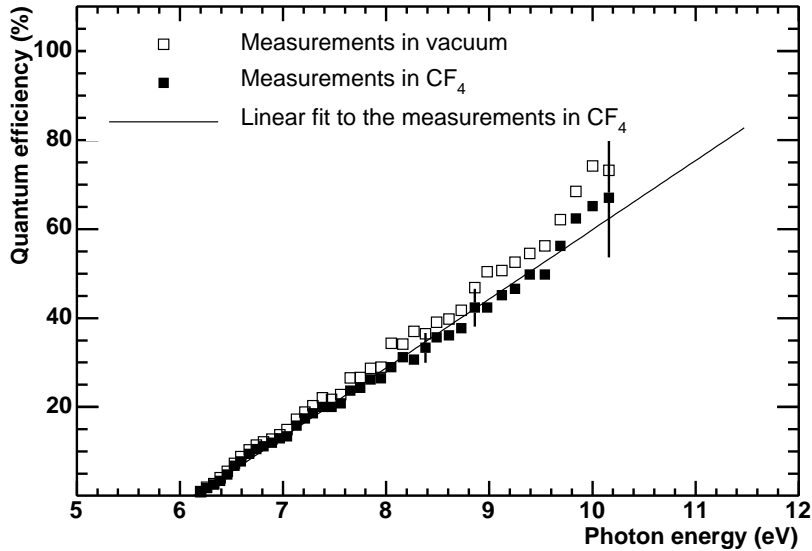}
    \caption{Absolute quantum efficiency of CsI in vacuum and $CF_{4}$ over
the bandwidth 6.2 - 10.3 eV.}
    \label{fig:qe_ev}
\end{figure}

Fig.~\ref{fig:qe_ev} shows the CsI absolute quantum efficiency in
vacuum and $CF_{4}$ plotted as a function of the photon energy.
(Plotting the quantum efficiency as a function of photon energy
has the advantage that the figure of merit N$_{0}$ is simply given
by the area under the points). The present results are in very
good agreement with those of ref \cite{qe11} which covered the
range 6-8.3 eV.

The range of the measurements shown in
Figs.~\ref{fig:photocurrent} and~\ref{fig:qe_ev} (120 - 200~nm, or
6.2 -10.3~eV) was limited by the light intensity of the deuterium
UV lamp at $\sim$ 10.3~eV. Over this bandwidth we derive a figure
of merit N$_0$ of 459 cm$^{-1}$ (414 cm$^{-1}$) or an average QE
value of 31\% (28\%) in vacuum (CF$_4$). However the useful range
of UV photons in the HBD extends up to the CF$_4$ cut-off which is
at ~11.5~eV. Extrapolating the data of Fig.~\ref{fig:qe_ev} to
11.5 eV under the assumption of a linear dependence of the quantum
efficiency vs. photon energy gives a figure of merit N$_{0}$= 822
cm$^{-1}$ or an average quantum efficiency of 55\% in CF$_4$. For
a 50 cm long radiator this N$_{0}$ value would correspond to
$\sim$ 35 photoelectrons (p.e.) (after taking into account the
losses due to the entrance mesh and the holes of the top GEM).
This unprecedented value of N$_{0}$ is a direct consequence of the
large bandwidth of CF$_4$ in the present windowless
configuration\footnote{A direct measurement of N$_{0}$ during the
beam test at KEK was not possible due to the large UV absorption
in the available gas system. We observed only 6-10 p.e. per
electron trigger. This number is however consistent with the
expected number of $\sim$ 35 p.e. after correcting it for the
measured absorption in the gas.}.

\section{Discharge probability and saturation effect.}

Stability of operation and absence of discharges in the presence
of heavily ionizing particles is crucial for the operation of the
HBD. The $^{241}$Am source was used to simulate heavily ionizing
particles under laboratory conditions. In our earlier
paper~\cite{NIM1} the discharge probability was measured in small
GEMs. In Ar/CO$_2$ it was found to increase sharply at a gas gain
of $\sim$2$\times$10$^4$ when the total charge approached the
Raether limit of 10$^8$~\cite{raeth} whereas in CF$_4$ no such
sharp increase was found. On the contrary the onset of the
discharges was spread over a broad gain range (see Fig. 4
in~\cite{NIM1}). The robustness of CF$_4$ against discharges
compared to Ar/CO$_2$ was attributed to the charge saturation
observed in CF$_4$ at the level of  ~2$\times$10$^7$ e (i.e. below
the Raether limit). However, all the measurements in CF$_4$ were
done with 30$\times$30 mm$^2$ GEMs. Similar measurements on large GEMs
could not be done since the first spark usually destroyed one of
the GEMs due to the large energy stored in the large capacitance.

\begin{figure}[htbp]
  \vspace{-10mm}
  \centering
  \includegraphics[keepaspectratio=true, width = 9cm]{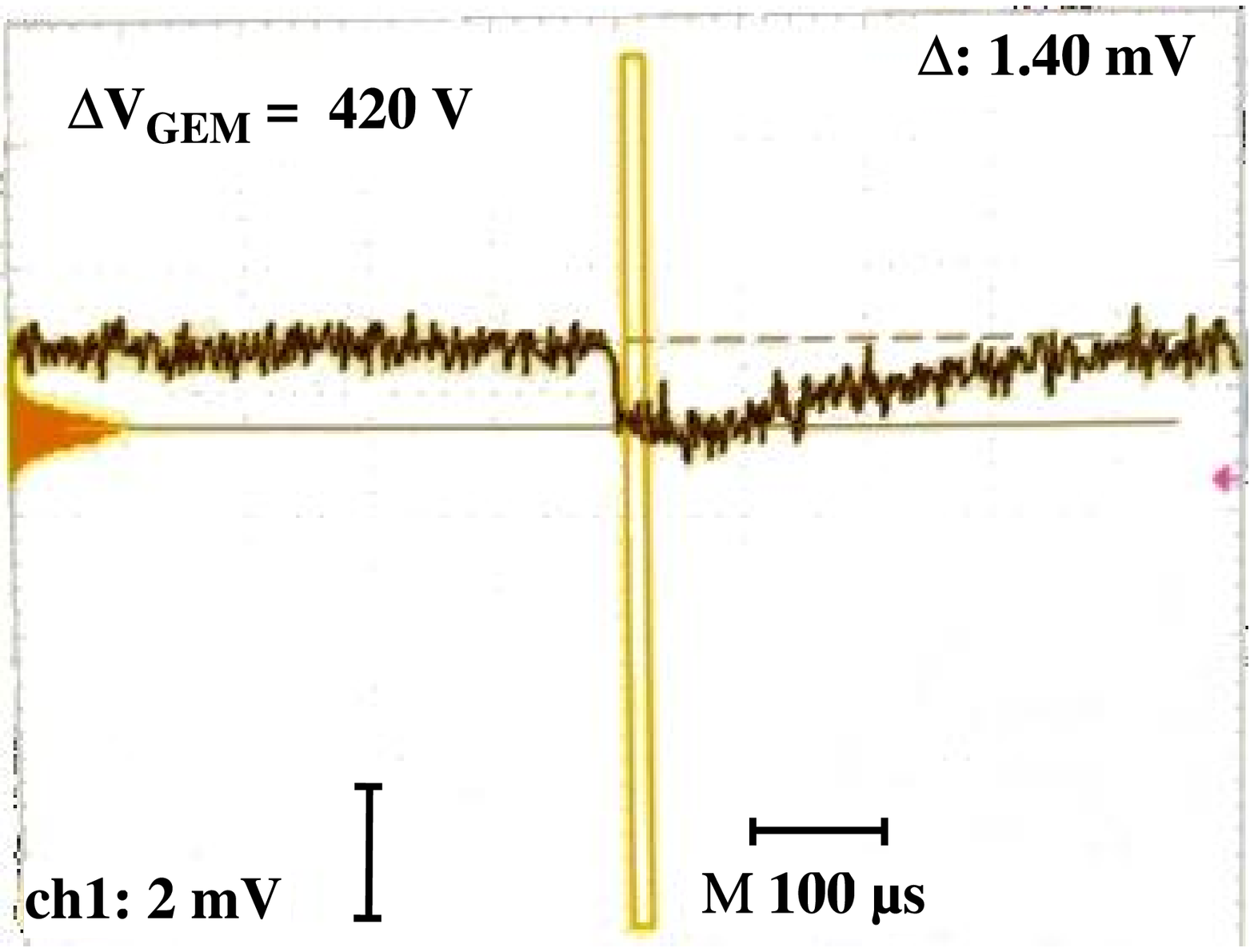}
  \hspace{-2cm}
  \vspace{0mm}
  \includegraphics[keepaspectratio=true, width = 9cm]{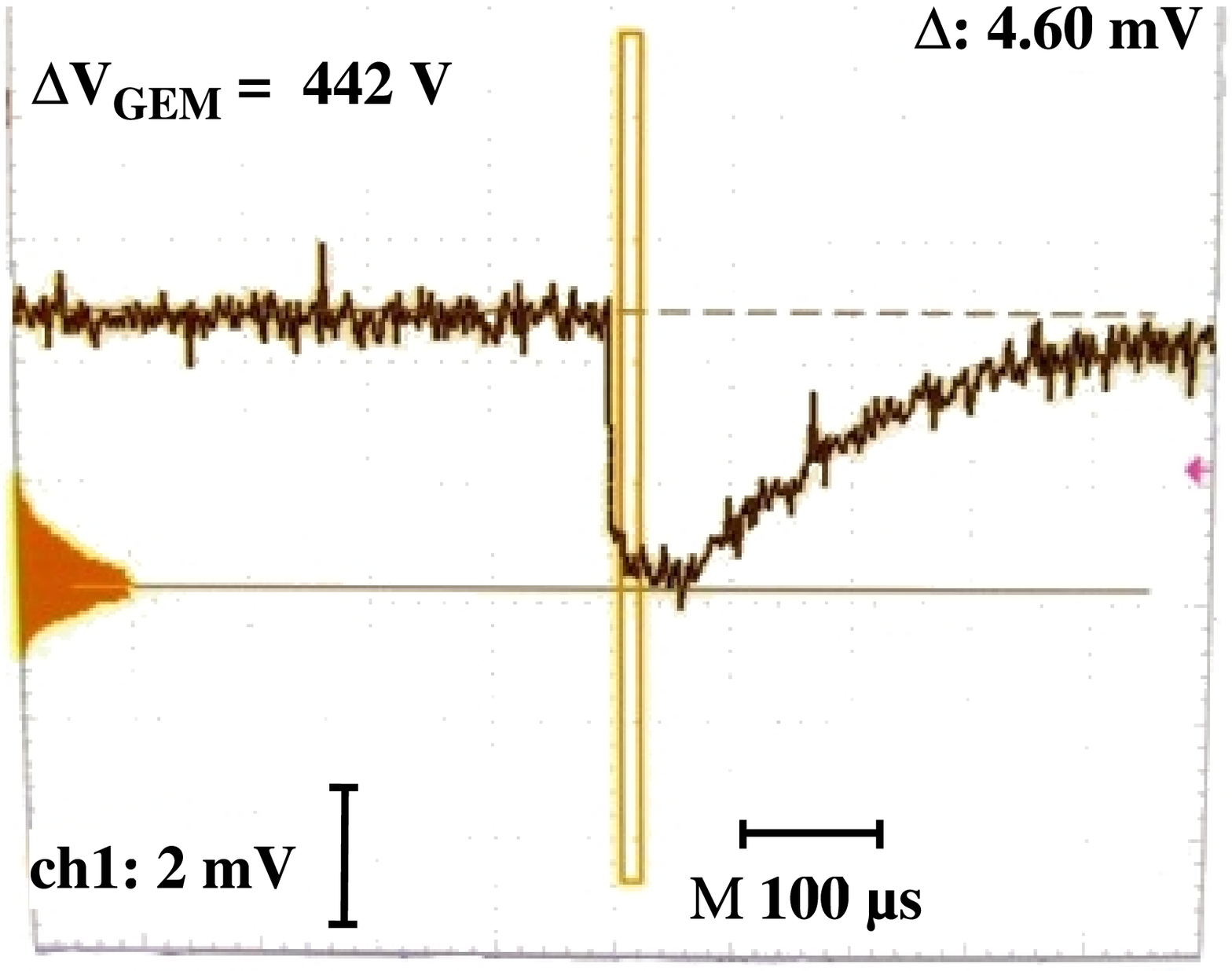}
  \hspace{-2cm}
  \vspace{10mm}
  \includegraphics[keepaspectratio=true, width = 9cm]{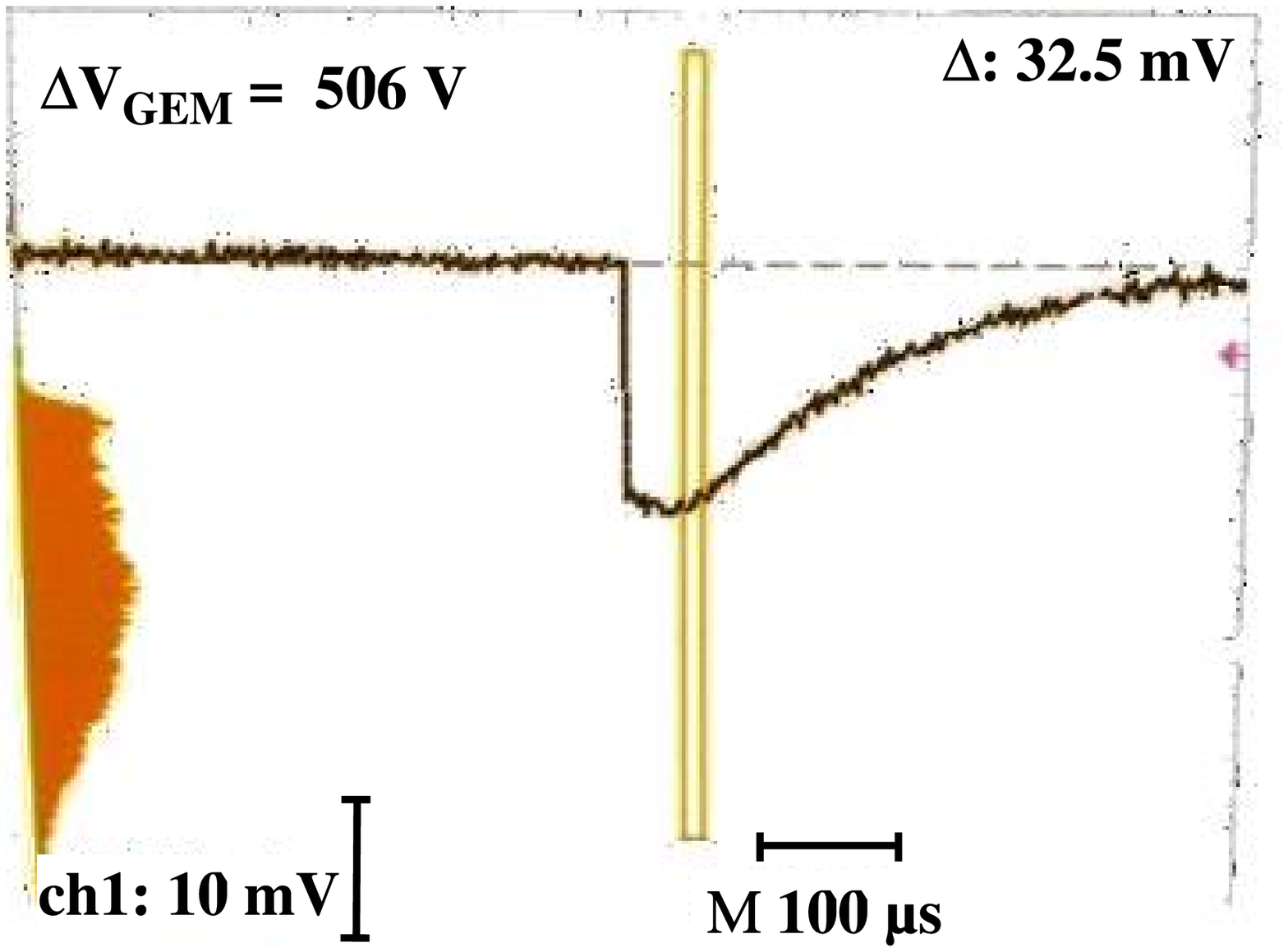}
  \caption{Examples of signals and pulse-height distributions (shown on the
   left side of the figures) measured without preamplifier. The detector was
   irradiated with $\alpha$-particles
  from the $^{241}$Am source. Figures a,b and c correspond to different
  voltages across the GEMs. The pulse-height signal is indicated
  in the top-right corner in mV.}
    \label{fig:dis2}
\end{figure}

\begin{figure}[ht]
    \centering
    \includegraphics[keepaspectratio=true, width = 12cm]{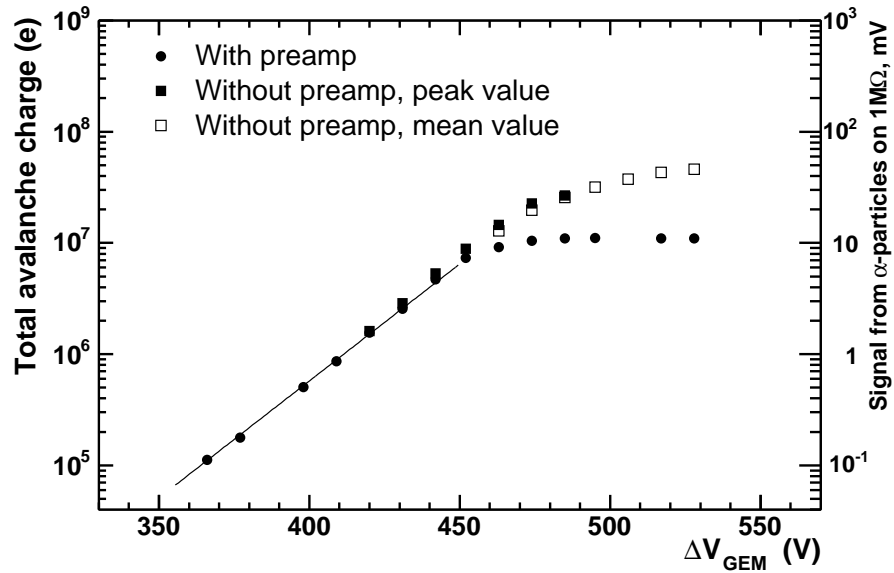}
    \caption{Pulse height of the signal from $\alpha$-particles measured with and
    without preamplifier as a function of GEM voltage. For the measurement with
    the preamplifier only mean values are plotted. For the measurement without the
    preamplifier both mean and peak values are plotted (solid and open
    squares respectively).}
    \label{fig:dis1}
\end{figure}

In the present work we used HV segmented 100$\times$100 mm$^2$
GEMs to repeat the discharge probability studies and revisit the
charge saturation effect in a more careful way. In particular, at
high values of the total charge, when the pre-amplifier is close
to saturation, the output signal is already reduced compared to
its real value. We therefore repeated the measurements of the
total charge produced by alpha-particles as a function of the GEM
voltage without any pre-amplifier, with the central pad directly
connected to the 1~M$\Omega$ input of an oscilloscope through a 1m
coaxial cable. In this case the pulse height observed at the scope
is determined by the ratio of the charge induced in the pad and
the total capacitance of the pad including the capacitance of the
cable. Examples of the pulses and pulse-height distributions
measured with this method are shown in Fig.~\ref{fig:dis2}. In the
figures the screen images of the oscilloscope are shown together
with the pulse height distribution on the left side of the
figures. The three figures correspond to 3 different voltages on
the GEMs. The risetime of the pulses is determined by the
induction of charge by the moving electrons and the decay time is
determined by the RC constant of the readout chain, i.e. the
product of the pad capacitance and the impedance of the scope
(1~M$\Omega$).  In Fig.~\ref{fig:dis1} we  compare the signal
measured without pre-amplifier at the 1~M$\Omega$ input of the
oscilloscope and the measurement with pre-amplifier. Both
measurements were performed  under identical conditions and for
this purpose the pre-amplifier was calibrated in units of input
charge. In order to obtain the relation beween the signal measured
without pre-amplifier and the input charge the results were
normalized in the range $\Delta$V$_{GEM}$ = 420 - 440~V where both
measurements could be performed and the amplifier was still far
from saturation. At $\Delta$V$_{GEM} >$ 490~V the pulse-height
resolution deteriorates considerably and we therefore plot the
mean value (instead of the peak value) of the pulse-height
distribution.

Fig.~\ref{fig:dis1} shows clearly that the signal from
alpha-particles deviates from the exponential dependence when the
total charge exceeds the value of 10$^7$ electrons.  The signal measured with
the pre-amplifier is saturated at the level of ~10$^7$~e (in the
measurements described in~\cite{NIM1} a pre-amplifier  with higher
saturation level was used). The signal measured without the pre-amplifier
saturates at about 4$\times$10$^7$~e. The bottom panel of Fig.~\ref{fig:dis2}
was obtained at $\Delta$V$_{GEM}$ = 506~V where according to Fig.~\ref{fig:dis1}
the saturation effect is already quite pronounced (the mean signal is suppressed by
almost a factor of 3 with respect to the expected exponential behaviour)
establishing a clear correlation between the saturation effect and the deterioration
of the pulse height resolution.

\begin{figure}[ht]
    \centering
    \includegraphics[keepaspectratio=true, width = 12cm]{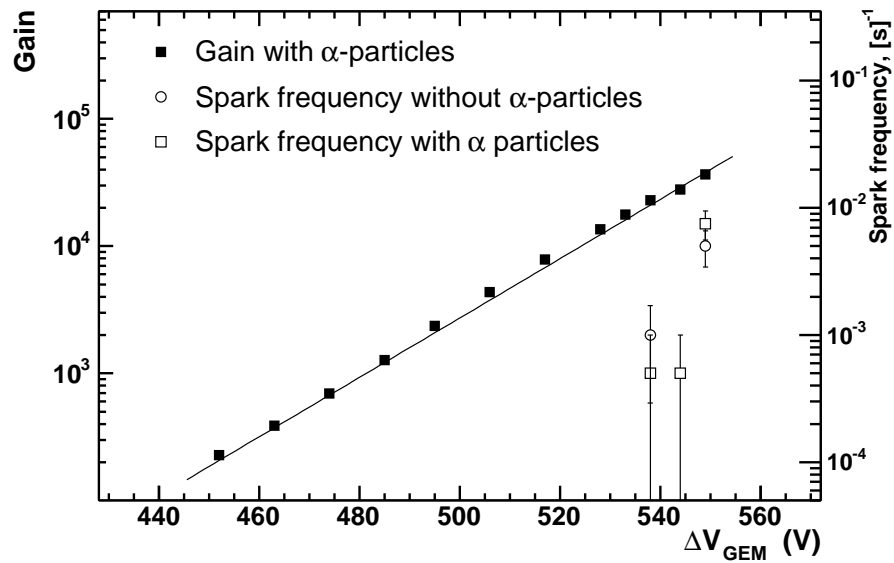}
    \caption{Spark frequency and detector gain as a function of voltage across
    the GEM with and without $\alpha$-particle irradiation.}
    \label{fig:dis3}
\end{figure}

As already mentioned the saturation of the signal strongly
suppresses the probability of the discharges provoked by heavily
ionizing particles. This suppression was also observed in the
measurements of the discharge limits performed with the segmented
large triple-GEM detector. The results of these measurements are
shown in Fig.~\ref{fig:dis3}. Two measurements are presented in
the figure, the measurement in the presence of alpha-particles and
the measurement without alpha-particles. In both cases the gain
was monitored at each voltage with the $^{55}$Fe source. The
dependence of the gain on voltage is presented in the figure
together with the spark frequency as a function of the GEM
voltage. The duration of each measurement was $\sim$ 2000~s, i.e.
the maximum number of sparks counted in the highest point was
about 20. The present detector showed negligible spark
probability at gains up to  $\sim 2\times10^4$.
The results demonstrate that the discharge limit does
not depend on the presence of alpha-particles in the sensitive
volume of the detector. Rather it seems that only local defects in
the GEMs are responsible for the discharges and limit the gain of
the device.

During the series of tests the detector experienced a total number
of 40 sparks but no sign of any degradation was observed in its
performance and all three GEMs were still fully operational at the
end of the test.


\section{Aging studies.}

In the present  HBD two possible sources of long term detector degradation
under irradiation can be identified: aging of the CsI photocathode
due to ion back-flow  and possible chemical reactions with active
radicals formed in electron-ion avalanches in CF$_4$, and aging of
the GEMs due to etching of metal or insulator surfaces and/or
polymerization of pollutants from avalanches in CF$_4$ at
the metal or insulator surfaces.  Both processes depend on the
total charge flowing through the detector which is the product of
the photo-electron current collected into the holes of the first
GEM and the  total gain of the triple-GEM system.

Assuming operation at a gain of $\sim 10^4$ we conservatively
estimate the total charge flow through the HBD in
normal PHENIX operation to be 10-20~$\mu$C/cm$^2$/year. Thus we
decided to perform an aging test of the detector module up to an
accumulated total charge of 100 to 200~$\mu$C/cm$^2$ to represent
approximately ten years of HBD operation in PHENIX.

The aging tests were performed so as to decouple the degradation
of the photocathode from the deterioration of GEMs. We used two
sources of radiation: The UV Hg-lamp and the $^{55}$Fe X-ray
source. The UV lamp was used for continuous irradiation of the
detector under test and the current to the PCB was monitored.
Every few hours the UV irradiation was stopped and
the $^{55}$Fe source was inserted into the detector for a short
gain monitoring allowing to assess the gain stability of the
triple-GEM. Once the gain is known we can then infer the stability
of the CsI photocathode. During UV irradiation the electric field
in the drift gap was kept at zero whereas during the X-ray
irradiation the drift field was set to $\sim$ 1~kV/cm.

\begin{figure}[htp]
    \hspace{-10mm}
  \begin{minipage}[t]{80mm}
    \includegraphics[keepaspectratio=true, width=9cm]{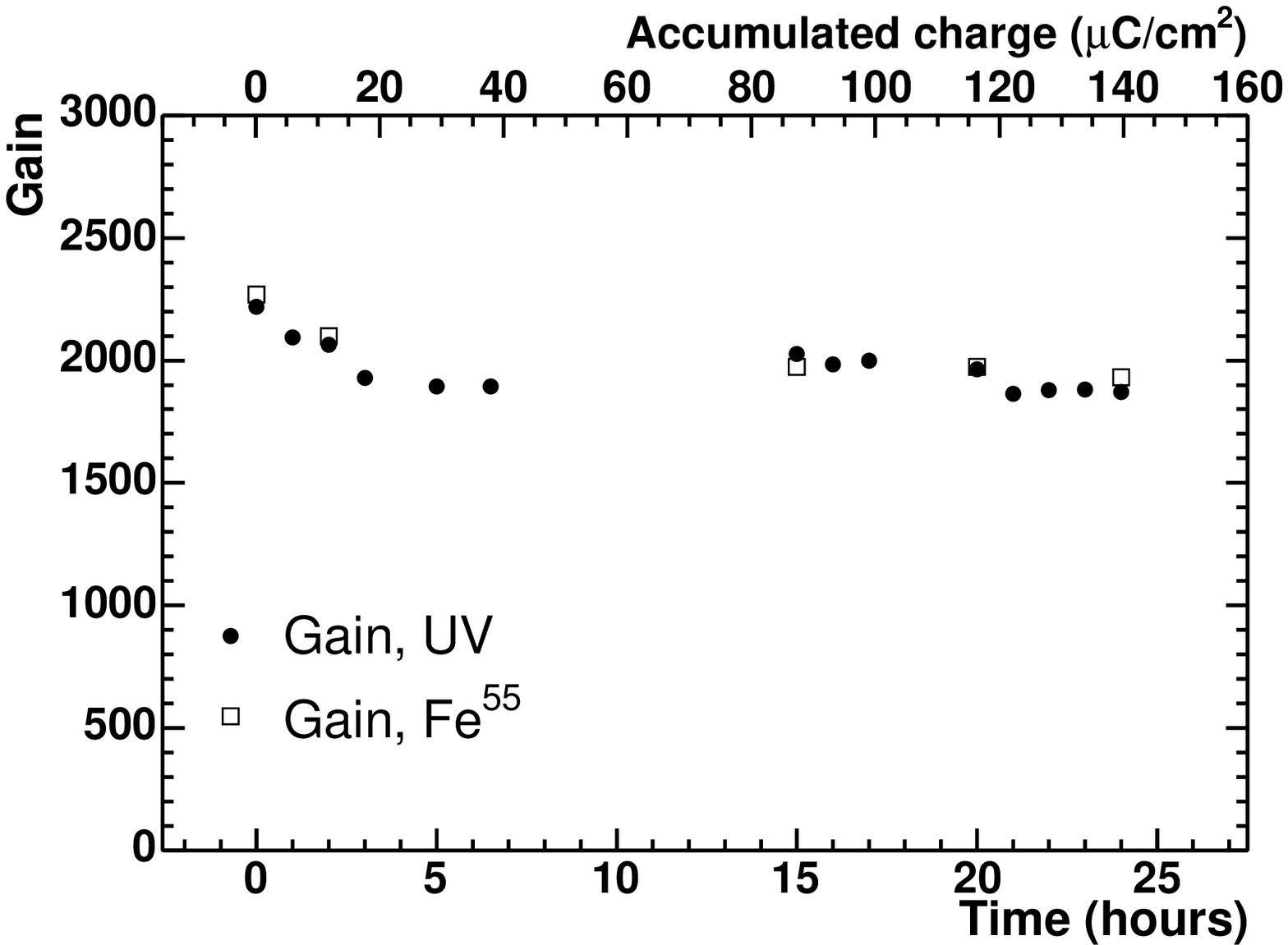}
  \end{minipage}
  \hspace{5mm}
  \begin{minipage}[t]{80mm}
    \includegraphics[keepaspectratio=true, width=9cm]{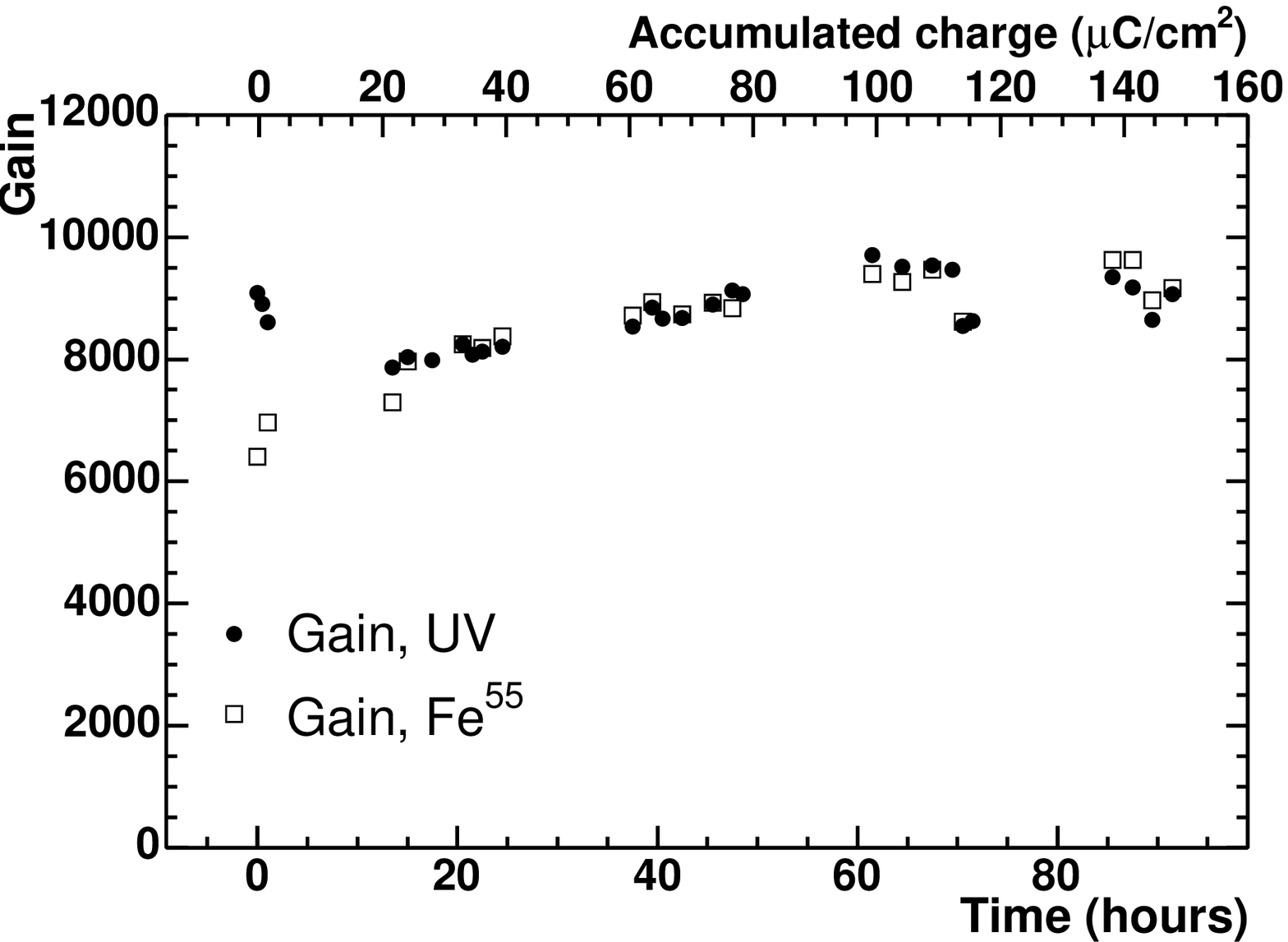}
  \end{minipage}
  \caption{Results of aging tests with 3$\times$3~cm$^2$ (left panel) and 10$\times$10~cm$^2$ (right panel)  triple-GEM detector
           with a CsI reflective photocathode. Open squares corresponds to the measurements with
           $^{55}$Fe, solid circles represents the measurements with UV irradiation.  }
  \label{fig:age12}
\end{figure}

In Fig.~\ref{fig:age12} the results of two aging tests are shown.
The first test (Fig.~\ref{fig:age12}, left) was performed with a
small triple-GEM  detector. The test took about 1 day and the
total charge accumulated was $\sim$ 140~$\mu$C/cm$^2$.
It is seen  that the gain derived from the measurements
of the current during UV-irradiation and the gain measured from
$^{55}$Fe irradiation are very close to each other. After an
initial gain drop of $\sim$ 10\% in the first 20 hours, the
performance was stable within ~2\%.

The second aging test was performed with a large segmented
triple-GEM set. This test was done at a lower rate than the first
one in that the detector accumulated a total charge of $\sim$
150~$\mu$C/cm$^2$ in 4 days rather than in 1 day. As seen in
the right panel of Fig.~\ref{fig:age12}, the gain variations
during the test did not exceed 20\% and during the second half of
the test they were even $\leq$~5\%.  In the first 10 hours of
irradiation the gain derived from UV irradiation and the one
determined from the measurements with the $^{55}$Fe source
differed by $\sim$~20\%. This result indicates that the
photocathode efficiency or collection efficiency of the
photo-electrons into the first GEM holes was higher during that
period. After the first 10 hours both gains converged to the same
value and  followed a very similar dependence. The performance of
both small and large GEM sets during the initial phases of the
aging tests including the gradual increase of the gain in the
second test is not yet understood\footnote{We have repeatedly
observed gain variations during the initial operation of a triple-GEM
over periods of time ranging from a few hours to a few days
before stable operation was reached. The origin of these
instabilities is unclear and will be the subject of further
studies.}. However as a result of these tests we conclude that
both the photocathode efficiency and the GEM gain do not exhibit
any dramatic change which can be interpreted as aging degradation
of the detector.

\section{Summary and conclusions}
An HBD is being proposed for an upgrade of the PHENIX experiment
at RHIC. The HBD is a windowless \v{C}erenkov detector, operated
with pure CF$_4$ in a proximity focus configuration. The
detector consists of a 50 cm long radiator directly coupled to a
triple-GEM detector which has a CsI photocathode evaporated on the
top face of the first GEM foil and a pad readout at the bottom of
the GEM stack.  We have studied the basic parameters which
determine the HBD performance. In particular, we have presented
results on the device response to mip's and to electrons. Large
hadron rejection factors, well in excess of 100, can be achieved
while preserving an electron detection efficiency larger than
90\%.
Extrapolating the quantum efficiency of CsI from the measured
range 6-10.3 eV to the expected operational bandwidth of the
device (6-11.5 eV) gives a figure of merit N$_0$=822 cm$^{-1}$ and
a very large number of photoelectrons of $\sim$~36 over a 50 cm
long radiator. The charge saturation effect occurring in CF$_4$
when the total charge in the avalanche reaches 4$\times$10$^7$~e
makes the HBD relatively robust against discharges.  Our
measurements show that the limit of stability is actually dictated
by the quality of the GEM foils rather than by the presence of
highly ionizing particles. Very stable operation can be achieved
at gains up to 10$^4$ with 10$\times$10 cm$^2$ GEMs segmented
into four HV segments\footnote{Very stable operation at gains of 10$^4$ was also
obtained with larger GEM foils of 23$\times$24 cm$^2$ that we have
tested recently.}. Aging studies of the GEM foils as well as the
CsI photocathode reveal that there is no significant deterioration
of the detector for irradiation levels corresponding to $\sim$~10
years of normal PHENIX operation at RHIC.
Recently, a test was carried out of a triple-GEM
detector operated with pure CF$_4$ and located inside the PHENIX
central arm spectrometer at a distance of 50 cm from the collision
point  \cite{phenix-test}. The detector performed smoothly in the presence of Au+Au
collisions exhibiting no discharges or gain instabilities. The
measurements presented here and the beam test results demonstrate the
validity of the proposed HBD concept and pave the way for the
incorporation of such a detector in the PHENIX experiment.

\section*{Acknowledgements}
\vspace{-5mm} We thank F.Sauli, A.Breskin, R.Chechik and M.Klin
for their invaluable help and very useful discussions. This work
was partially supported by the Israel Science Foundation, the
MINERVA Foundation, the Nella and Leon Benoziyo Center for High
Energy Physics Research and the US Department of Energy under
Contract No. DE-AC0298CH10886.

\end{document}